\documentclass[twocolumn,preprintnumbers,amsmath,amssymb,superscriptaddress]{revtex4-2}
\usepackage{eqnarray,amsmath,lipsum}
\usepackage{graphicx}% Include figure files
\usepackage{dcolumn}
\usepackage{natbib}

\usepackage{color}
\definecolor{red}{rgb}{1,0,0}

\definecolor{blue}{rgb}{0,0,1}

\definecolor{green}{rgb}{0,1,0}

\setcitestyle{square}

\begin{document}
\preprint{APS}

\author {Lucas Squillante}
\affiliation{S\~ao Paulo State University (Unesp), IGCE - Physics Department, Rio Claro - SP, Brazil}
\author{Luciano S. Ricco}
\affiliation{Science Institute, University of Iceland, Dunhagi-3, IS-107, Reykjavik, Iceland}
\author {Aniekan Magnus Ukpong}
\affiliation{Theoretical and Computational Condensed Matter and Materials Physics Group, School of Chemistry and Physics, University of KwaZulu-Natal, Pietermaritzburg, South Africa}
\affiliation{National Institute for Theoretical and Computational Sciences (NITheCS), KwaZulu-Natal, South Africa}
\author {Roberto E. Lagos-Monaco}
\affiliation{S\~ao Paulo State University (Unesp), IGCE - Physics Department, Rio Claro - SP, Brazil}
\author {Antonio C. Seridonio}
\affiliation{S\~ao Paulo State University (Unesp), Department of Physics and Chemistry, Ilha Solteira - SP, Brazil}
\author {Mariano de Souza}
\email{mariano.souza@unesp.br}
\affiliation{S\~ao Paulo State University (Unesp), IGCE - Physics Department, Rio Claro - SP, Brazil}

\title{Gr\"uneisen parameter as an entanglement compass and the breakdown of the Hellmann-Feynman theorem}

\begin{abstract}
The Gr\"uneisen ratio $\Gamma$, i.e., the singular part of the ratio of thermal expansion to the specific heat, has been broadly employed to explore both finite-$T$ and quantum critical points (QCPs). For a genuine quantum phase transition (QPT), thermal fluctuations are absent and thus the thermodynamic $\Gamma$ cannot be employed. We propose a quantum analogue to $\Gamma$ that computes entanglement as a function of a tuning parameter $\lambda$ and show that QPTs take place only for systems in which the ground-state energy depends on $\lambda$ non-linearly. Furthermore, we demonstrate the breakdown of the Hellmann-Feynman theorem in the thermodynamic limit at any QCP. We showcase our approach using the quantum 1D Ising model with transverse field and Kane's quantum computer. %Implications of our findings to quantum computing are presented.
The slowing down of the dynamics and thus the ``creation of mass'' close to any QCP/QPT is also discussed.
\end{abstract}

\maketitle
\vspace{1.0cm}
\date{\today}
%\section{Introduction}
The investigation of the thermodynamic response close to a finite-$T$ critical end point is a widely explored branch \cite{stanley}. A key physical quantity related to critical behaviour of matter is the correlation length $\xi$. As the temperature $T$ and tuning parameter approaches their critical values, $\xi \rightarrow \infty$ \cite{stanley}. This is because of the emergence of large fluctuations of the order parameter and thus intrinsic entropy accumulation in this regime \cite{zhu,mrb}. Such enhancement of the entropy in the vicinity of critical points can be used to attain giant caloric effects \cite{mrb}. Interestingly enough, critical phenomena can be observed in various fields, including complex biological systems, such as the brain and the cognition development \cite{Munoz2013,grigolini}, which has provided a new foundation for modern Thermodynamics.
%\section{Gr\"uneisen parameter for $T = 0$\,K}
Apart from $\xi$, an appropriate tool to explore critical phenomena is the so-called Gr\"uneisen ratio $\Gamma_g$ \cite{ejp,mrb}:
\begin{equation}
\Gamma_g = \frac{\alpha_g}{c_g} = -\frac{1}{T}\frac{\left(\frac{\partial^2 F}{\partial T \partial g}\right)}{\left(\frac{\partial^2 F}{\partial T^2}\right)_g} = -\frac{1}{T}\frac{\left(\frac{\partial S}{\partial g}\right)_T}{\left(\frac{\partial S}{\partial T}\right)_g} = \frac{1}{T}\left(\frac{\partial T}{\partial g}\right)_S,
\label{gamma}
\end{equation}
where $\alpha_g$ and $c_g$ are, respectively, the thermal expansivity and heat capacity, both at a constant value of a tuning parameter $g$; $F$ and $S$ are, respectively, the Boltzmann free energy and entropy. For the case in which $g$ is pressure, $\Gamma_g$ quantifies the barocaloric effect. Stress, magnetic, and electric field can be used as $g$, being, respectively, each of them associated with the elastic-, magnetic-, and electric-Gr\"uneisen parameters \cite{mrb,elastocaloric}. The peculiar behaviour of $\Gamma_g$ close to finite-$T$ critical points, namely an enhancement and sign-change upon varying $g$ is well-established in the literature \cite{griffiths,prl}. The same holds true for quantum critical points (QCPs) \emph{near} absolute zero \cite{unveiling,zhu}. In this context,  it is tempting to ask ``What is the expression for $\Gamma_g$ at exactly $T = 0$\,K?'' This is the key-question we have addressed in the present work. Note that as $T \rightarrow 0$\,K, both $\alpha_g$ and $c_g$ $\rightarrow 0$ because of the third-law of Thermodynamics \cite{stanley}, being clear that in such a case, $\Gamma_g$ is undetermined. At first glance, one could consider that as $T \rightarrow 0$\,K $\Rightarrow$ $\Gamma_g \rightarrow \infty$ as a mere consequence of the vanishing of the energy scale associated with the thermal energy, cf.\,Eq.\,\ref{gamma}. It turns out that at $T = 0$\,K, Eq.\,\ref{gamma} no longer holds true and the classical concept of caloric effects is no longer applicable. Regarding QCPs, the behaviour of $\Gamma_g$ has been investigated, in practice, at ultra-low temperatures since typical order parameter fluctuations associated with QCPs are accessible in the $T$ range of a few mK \cite{unveiling,zhu}. However, as pointed out previously, the expression for $\Gamma_g$ is no longer valid exactly at $T = 0$\,K preventing thus the analysis of a genuine QCP.

Here, we propose an analogous expression for $\Gamma_g$ which is universal, valid at absolute zero temperature, and might be relevant for quantum computing. To this end, we start considering a generalized Hamiltonian given by:
\begin{equation}
H = H_0 + H_1(h) + H_2(g),
\end{equation}
where $H_0$ is the unperturbed Hamiltonian, $H_1(h)$ and $H_2(g)$ are perturbed terms, whose eigenenergies $E_i$ depend on the tuning parameters $h$ and $g$, e.g., magnetic or electric fields. Also, our proposal covers the case of $h$ and $g$ being two competing energy scales. Since we are dealing with a quantum system, we propose, in analogy to $\alpha_g$, to consider the numerator of the first part of Eq.\,\ref{gamma} as $\frac{\partial^2E_0}{\partial h \partial g}$, where $E_0$ is the ground-state energy and; for the denominator, we propose, in analogy to $c_g$, to consider $-h\frac{d^2 E_0}{dh^2}$, following similar arguments presented in Ref.\,\cite{dobes}. The $g$ and $h$ derivatives of $E_0$ are related to the Hellmann-Feynman theorem, namely $(dE(\nu)/d\nu) = \langle\psi_n| \frac{dH(\nu)}{d\nu}|\psi_n\rangle$, where $\nu$ is a general tuning parameter and $\psi_n$ a single-particle wavefunction \cite{roscilde}.
Hence, upon connecting the dots, a mathematical expression for the Gr\"uneisen parameter that is valid at $T = 0$\,K reads:
\begin{equation}
\Gamma^{0\textmd{K}} = -\frac{\left(\frac{\partial^2 E_0}{\partial h\partial g}\right)}{h\left(\frac{\partial^2 E_0}{\partial h^2}\right)},
\label{gamma0K}
\end{equation}
whose dim is $[\textmd{g}]^{-1}$. Although the second and cross derivatives of $E_0$ with respect to tuning parameters have been employed to investigate quantum phase transitions, see, e.g., Ref.\,\cite{crossderivative}, to the best of our knowledge, a proper discussion about its genesis, i.e., its correspondence with the Gr\"uneisen parameter, was still lacking. In a real physical system, it is natural to expect that $E_0$ might be more sensitive with respect to either $h$ or $g$ depending on particular physical aspects of the investigated system. Upon considering that $E_0$ is changed as a result of the variation of $h$ and/or $g$, the quantum entropy of the system is affected. More precisely, changing $E_0$ lead to a variation of the von Neumann entropy $S_N = -\textmd{Tr}(\rho\ln{\rho})$ \cite{neumann}, where $\rho = \sum_j p_j|\psi_j\rangle\langle\psi_j|$ is the density matrix and $p_j$ is the probability of each state $|\psi_j\rangle$. This can be demonstrated upon making use of the average energy of the system $\langle H \rangle$ in a state $|\psi\rangle$ in terms of $\rho$, i.e., $\langle H \rangle = \textmd{Tr}(\rho H)$ \cite{sakurai}. Considering that the system lies in the ground-state, its average energy is $E_0$ and thus its variation due to a change of $h$ and $g$ means that $\rho$ is also changed. As a consequence, if $\rho$ is varied, $S_N$ is varied as well. Hence, we write $\Gamma^{0\textmd{K}}$ in terms of $S_N$ (see Eq.\,\ref{gamma}), namely:
\begin{equation}
\Gamma^{0\textmd{K}} = -\frac{\left(\frac{\partial S_N}{\partial g}\right)_{h}}{h\left(\frac{\partial S_N}{\partial h}\right)_g} = -\frac{\left(\frac{\partial \textmd{Tr}(\rho\ln{\rho})}{\partial g}\right)_{h}}{h\left(\frac{\partial \textmd{Tr}(\rho\ln{\rho})}{\partial h}\right)_g}.
\label{entropy}
\end{equation}
In our case, we have considered $S_N$ as the global entanglement, being defined as $S_N = -(\partial E_0/\partial h)$, in a direct analogy to the thermodynamic case \cite{stanley}. Since $S_N$ quantifies the entanglement of a quantum system \cite{wootters1}, the definition of $\Gamma^{0\textmd{K}}$ in Eq.\,\ref{entropy} can be regarded as an ``\emph{entanglement compass}'' to explore QCPs upon varying $h$ and $g$.

The detection of genuine quantum phase transitions (QPTs) and QCPs is key in the context of quantum computing, since close to both QPTs and QCPs, the system's states are highly entangled \cite{fazio}. This is corroborated by the so-called concurrence, which quantifies the degree of entanglement between two qubits \cite{wootters1,wootters2,fazio}. Hence, $\Gamma^{0\textmd{K}}$ can also compute the concurrence between two qubits upon varying $h$ and/or $g$, cf.\,Eq.\,\ref{entropy}. In Ref.\,\cite{fazio}, it is discussed the field switching of the entanglement for the canonical 1D Ising model under transverse field (IMTF), demonstrating that close to the QCP the entanglement is strongly magnetic-field-dependent. Hence, we consider that $\Gamma^{0\textmd{K}}$ is the appropriate physical quantity to quantify the latter.
It turns out that applying $\Gamma^{0\textmd{K}}$ (Eq.\,\ref{gamma0K}) to the $S = 3/2$ Hamiltonian with spin-orbit interaction for the case of a Brillouin-like paramagnet \cite{ney,prb}, hydrogen atom under external magnetic field \cite{feynman}, Kondo lattice \cite{kondo}, and Majorana nanowires \cite{marra} leads to $\Gamma^{0\textmd{K}} \rightarrow \infty$ for $h, g \rightarrow 0$, which is the trivial case, i.e., no critical features show up for finite $h$ or $g$. Also, their cross and second derivatives of $E_0$ in respect to $h$ and $g$ do not present any divergence, which suggests that QCPs/QPTs are absent for these systems.

\begin{figure}[t]
\centering
\includegraphics[width=\columnwidth]{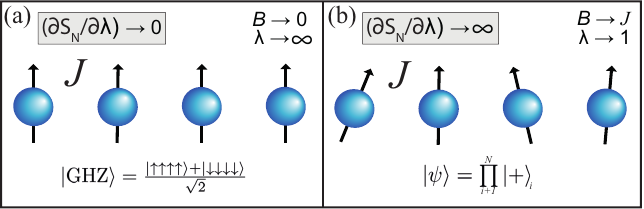}
\caption{\footnotesize Schematic representation of the here-proposed optimization of quantum information processing for the 1D IMTF. a) For $B \rightarrow 0$ $\Rightarrow$ $\lambda \rightarrow \infty$, the ferromagnetic phase state is the Greenberger-Horne-Zeilinger-like state (GHZ) \cite{asadian}, i.e., a superposition of $\uparrow$ and $\downarrow$ states. The von Neumann entropy $S_N$ is insensitive to $\lambda$ changes. b) Upon applying the transverse field, the system is brought to the quantum paramagnetic phase, i.e., $\lambda \rightarrow 1$, being the system state defined as the product of each eigenstate  and $|+\rangle = 1/\sqrt{2}(|\uparrow\rangle + |\downarrow\rangle)$ \cite{soh}, which corresponds to the simplest entangled state. In this regime, $S_N$ dramatically changes upon varying $\lambda$. For a real system, one has to properly set the initial state (panel a)) before applying a transverse $B$ to achieve the here discussed optimal conditions for quantum computing.}
\label{Fig-2}
\end{figure}

Regarding the case in which $E_0$ is non-linear with respect to the tuning parameter, $\Gamma^{0\textmd{K}}$ can be employed to probe QCPs and QPTs. This is the case of the 1D IMTF, whose Hamiltonian is given by \cite{sachdev,pfeuty}:
\begin{equation}
H = -B\sum_i S_i^x - J\sum_iS_i^zS_{i+1}^z,
\label{transverse}
\end{equation}
where $B$ is the modulus of the transverse magnetic field, $S_i^x$ is the spin operator at the $i$ site along the $x$-axis, $J$ is the exchange coupling constant between nearest neighbours magnetic moments, and $S_i^z$ and $S_{i+1}^z$ are the spin operators along the $z$-axis at sites $i$ and $i+1$, respectively. Before computing $\Gamma^{0\textmd{K}}$, $E_0$ must be obtained for this case. At this point, it is worth mentioning that $\Gamma$, but not $\Gamma^{0\textmd{K}}$, was already computed for this case in the vicinity of the QCP showing a divergent-like behaviour when the magnetic energy matched $J$, i.e., for $\lambda = J/B = 1$ \cite{qsi}. To compute $E_0$ for Eq.\,\ref{transverse}, a textbook Jordan-Wigner transformation can be employed and the Hamiltonian is diagonalized, so that \cite{bog}:
\begin{equation}
H = B\sum_k \Lambda_k\eta_k^{\dagger}\eta_k - \frac{B}{2}\sum_k\Lambda_k,
\end{equation}
where $\Lambda_k = 2\sqrt{1 + \lambda^2 - 2\lambda\cos{(k)}}$ and $\eta_k^{\dagger}$ and $\eta_k$ are, respectively, the creation and annihilation operators, being then $E_0$ per site given by $E_0 = -(B/\pi)\sum_k\Lambda_k$ \cite{pfeuty}, which in turn can be rewritten in terms of an elliptical integral in the thermodynamic limit \cite{sebastian}:
\begin{equation}
E_0 = -\frac{B}{\pi}\int_0^{\pi}\Lambda_k dk = -\frac{4B\sqrt{\lambda^2 + 2\lambda +1}\textmd{Ell}[-\frac{4\lambda}{(\lambda-1)^2}]}{\pi\sqrt{\frac{\lambda^2 + 2\lambda+1}{(\lambda-1)^2}}},
\end{equation}
\begin{figure}[t]
\centering
\includegraphics[width=1.03\columnwidth]{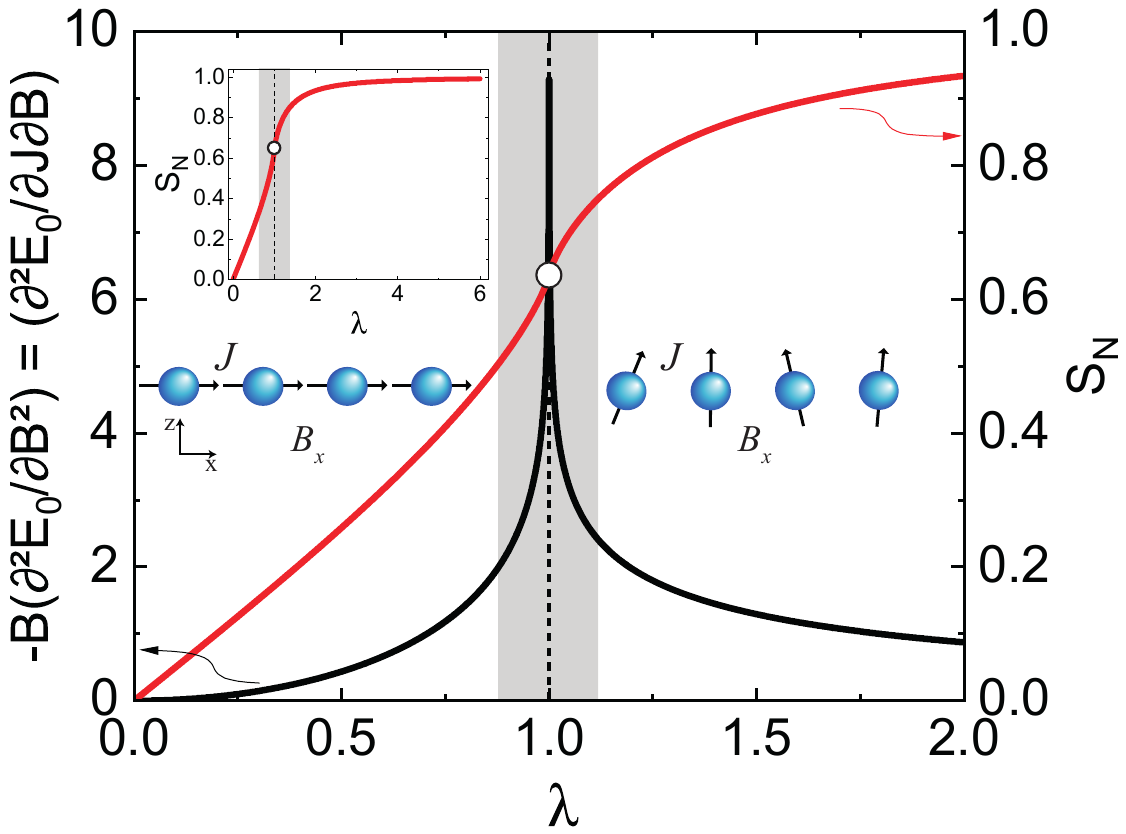}
\caption{\footnotesize Cross ($\partial {E_0}^2/\partial J\partial B$) and second $-B(\partial^2 E_0/\partial B^2$) derivatives of the ground-state energy $E_0$ (black line) and the von Neumann entropy $S_N$ as a function of $\lambda = J/B$. The gray region indicates the $\lambda$ range where both $E_0$ derivatives and $S_N$ vary more expressively with $\lambda$. The spin configurations depict the ferromagnetic phase for $\lambda < 1$ and the quantum paramagnet close to $\lambda = 1$. Inset: $S_N$ \emph{versus} $\lambda$ for large values of $\lambda$, i.e., for $S_N \rightarrow$ 1, the so-called Greenberger-Horne-Zeilinger-like state is achieved \cite{asadian}. The white bullet depicted in $S_N$ \emph{versus} $\lambda$ both in the main panel and inset indicates $S_N \rightarrow \infty$ for $\lambda = 1$.}
\label{Fig-1}
\end{figure}
where $\textmd{Ell}[-4\lambda/(\lambda-1)^2]$ is the so-called complete elliptic integral of the second kind \cite{gradshteyn}. Hence, upon choosing $h = B$ and $g = J$, $\Gamma^{0\textmd{K}}$ can be computed exactly.
It turns out that for the 1D IMTF both $E_0$ derivatives in the numerator and denominator of Eq.\,\ref{gamma0K} are exactly the same, which is in contrast with the thermodynamic case, cf.\,Eq.\,\ref{gamma}. This is merely a consequence of the fact that $E_0$ has the same dependence on $J$ and $B$. Figure\,\ref{Fig-1} depicts the $\lambda$ derivative of $E_0$. Remarkably, for $\lambda = 1$, $\frac{\partial^2 E_0}{\partial J\partial B}$ and $-B\frac{\partial^2 E_0}{\partial B^2}$ $\rightarrow$ $\infty$, a fingerprint of a QCP governing the change from a ferromagnetic to a quantum paramagnetic phase \cite{qsi}. Also, the result of $S_N$ as a function of $\lambda$ is depicted in Fig.\,\ref{Fig-1}. An enhancement in $S_N$ is observed for $\lambda = 1$ analogously to the thermodynamic case of a phase transition, which in turn can be discussed in terms of either a first- or second-order  \cite{stanley}. It is notorious that for $\lambda = 1$, $S_N$ is very sensitive to subtle changes of the tuning parameter. Hence, the system's entanglement in that region is enhanced analogously to the case of caloric effects close to finite-$T$ critical points \cite{mrb}. It is clear that in both cases the adiabatic character is crucial \cite{albash}, considering that real applications to quantum computing takes place at finite $T$. Note that we propose that upon considering the tuning parameter-dependence of $E_0$, $S_N$ can be computed straightforwardly by making the first derivative of $E_0$ with respect to the tuning parameter, i.e., a reduced density matrix is not needed in our case to compute global entanglement as it is done in the frame of the Meyer-Wallach approach \cite{meyer}. This is corroborated by the fact that our result for $S_N$ \emph{versus} $\lambda$ is in perfect agreement with literature results employing the Meyer-Wallach approach to compute global entanglement in the thermodynamic limit, cf.\,Fig.\,1 of Ref.\,\cite{asadian}. The enhancement of $(\partial S_N/\partial\lambda)$ in the immediate vicinity of the QCP for the 1D IMTF was already reported \cite{asadian}. However, our proposal of $\Gamma^{0\textmd{K}}$ universalizes the behaviour of $S_N$ in terms of $E_0$ changes due to the tuning of either $h$ or $g$ for any quantum system, including the 2D and 3D versions of the IMTF \cite{asadian}. Such an enhancement of $(\partial S_N/\partial\lambda)$ close to the QCP is associated with a violation of the Hellmann-Feynman theorem \cite{roscilde}, which in turn reflects the emergence of low-energy spin wave excitations for $\lambda \simeq 1$ \cite{sethna}. As it is known in the literature, $S_N$ scales logarithmically with $N$ at the QCP and diverges for an infinite chain ($N \rightarrow \infty$) \cite{fazio,kitaev}, making $(\partial S_N/\partial\lambda)$ to be enhanced in the vicinity of the QCP and to diverge right at the QCP. Hence, we unprecedentedly recognize that $S_N = -(\partial E_0/\partial \lambda)$ is connected to the Hellmann-Feynman theorem. The divergence of the first derivative of $E_0$ with respect to $\lambda$, i.e., $S_N$, at the QCP indicates a breakdown of the Hellmann-Feynman theorem for an infinite chain. Such a breakdown is not expected to take place for finite chains, cf.\,Refs.\,\cite{fazio,kitaev}. Figure\,\ref{Fig-2} depicts our proposal of a controlled tuning of $S_N$ for the 1D IMTF. For $B \rightarrow 0$ ($\lambda \rightarrow \infty$) $\Rightarrow$ $S_N \rightarrow 1$, cf.\,inset of Fig.\,\ref{Fig-1}, and thus $(\partial S_N/\partial \lambda) \rightarrow 0$ meaning that $S_N$ is insensitive to $B$ variations in this regime, corresponding to the Greenberger-Horne-Zeilinger-like (GHZ) state \cite{asadian}. For $B \rightarrow J$ ($\lambda \rightarrow 1$), a quantum paramagnetic phase is established, being $S_N$ dramatically affected by small changes of $B$. Hence, it is clear that in the quantum paramagnetic phase, i.e., in the regime $\lambda \simeq 1$, the system's entanglement can be easily manipulated upon varying the tuning parameter, which is $B$ in the case of the 1D IMTF.

\begin{figure}[t]
\centering
\includegraphics[width=0.89\columnwidth]{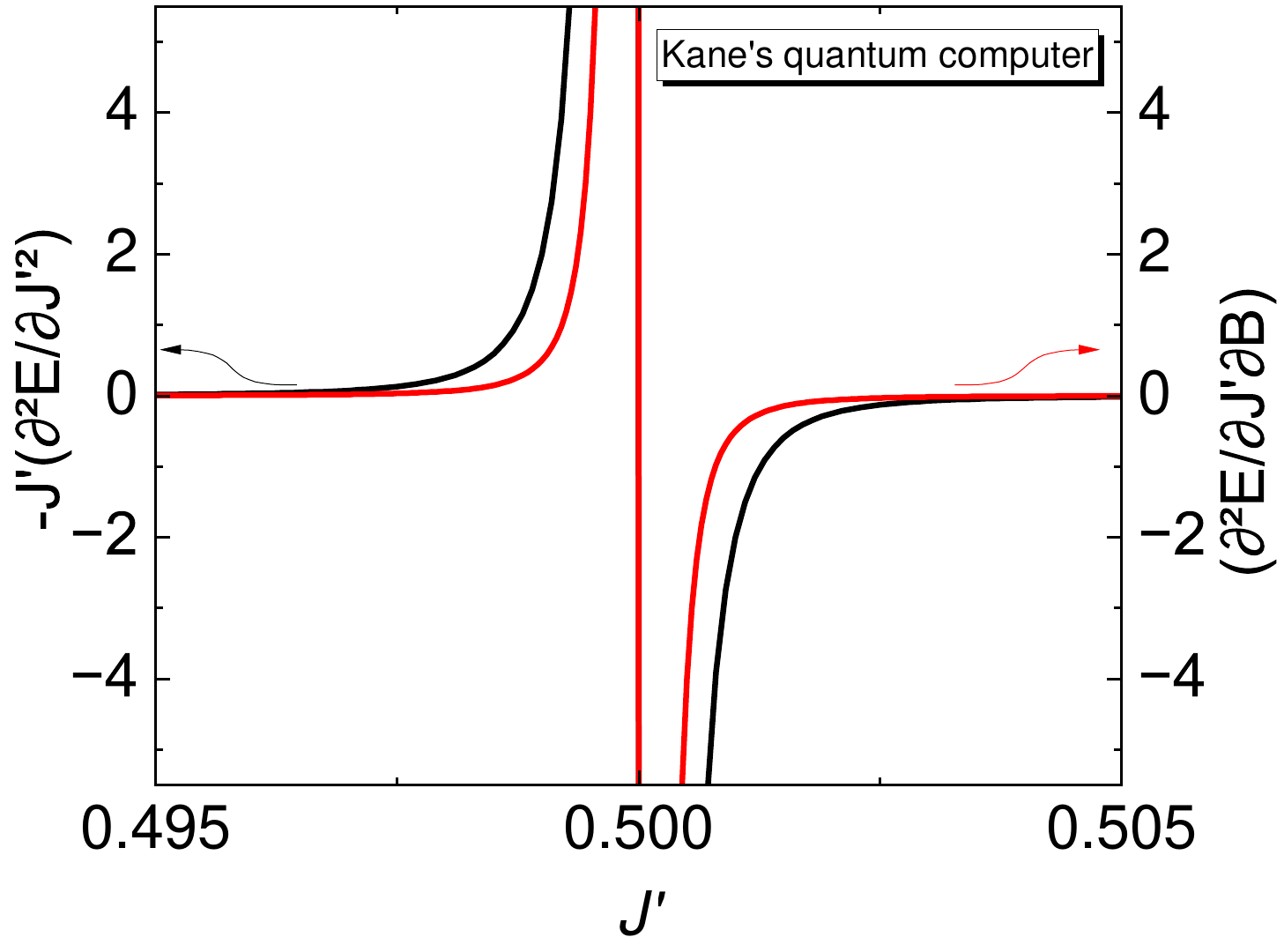}
\caption{\footnotesize Cross ($\partial E^2/\partial J'\partial B$) and second $-J'(\partial^2 E/\partial J'^2$) derivatives of Eq.\,\ref{energy} for Kane's quantum computer using $B = 1$\,T. Details in the main text.}
\label{Fig-3}
\end{figure}

It turns out that entanglement is a key ingredient for quantum computing \cite{steane}. Hence, the here proposed $\Gamma^{0\textmd{K}}$ can be considered as an \emph{entanglement compass} \cite{assad}, having direct implications for Kane's quantum computer. The latter is based on the Hamiltonian for two interacting nucleus-electron $^{31}$P atoms embedded in a Si matrix, namely \cite{kane}:
\begin{equation}
H = H(B) + A_1\sigma^{1n}\cdot\sigma^{2e} + A_2\sigma^{2n}\cdot\sigma^{2e}+ J'\sigma^{1e}\cdot\sigma^{2e},
\end{equation}
where $H(B)$ is the Zeeman interaction, $A_1$ and $A_2$ are, respectively, the hyperfine coupling constant between the first and second $^{31}$P nuclear and electronic magnetic moments, and $J'$ the coupling constant between electronic magnetic moments. Upon considering $A_1 = A_2 = A$, the energy $E$ difference between states $|10 - 01\rangle$ and $|10 + 01\rangle$ is given by \cite{kane}:
\begin{equation}
E = h\nu_{J'} = 2A^2\left(\frac{1}{\mu_B B - 2J'} - \frac{1}{\mu_B B}\right),
\label{energy}
\end{equation}
where $\nu_{J'}$ is the nuclear spin exchange frequency and $\mu_B$ the Bohr magneton. Thus, Eq.\,\ref{gamma0K} can be employed straightforwardly to compute $\Gamma^{0\textmd{K}}$ for Kane's quantum computer considering $h = J'$ and $g = B$. Figure\,\ref{Fig-3} shows that, analogously to the case of the 1D IMTF, when $\mu_B B = 2J'$, $S_N$ is dramatically enhanced, which in turn means that Kane's quantum computer achieves a condition in which information processing capacity is increased. Also, upon analyzing the results depicted in Fig.\,\ref{Fig-3}, it is remarkable that both derivatives, namely $-J'(\partial^2 E/\partial J'^2$) and $(\partial^2 E/\partial J'\partial B)$, are not exactly the same. This is because the $J'$ and $B$ dependencies of $E$ are distinct, cf.\,Eq.\,\ref{energy}. A sign-change of both derivatives can be observed for $J' = 0.5$ due to the distinct configurations between magnetic moments. It is clear that Kane's quantum computer is a bipartite system, i.e., two qubits are considered \cite{kane}. Hence, we can extend our analysis to the so-called multipartite systems, where $n$ qubits are considered and $\rho_n = \frac{1}{2^n}I_n$ \cite{buzek}, where $I_n$ is the identity operator. The latter describes $\rho$ for a chain of $n$ qubits, making it possible to calculate $S_N$ for any multipartite system where $\rho_n$ is known. By computing $S_N$ for a multipartite system, the derivatives incorporated in the here-proposed $\Gamma^{0\textmd{K}}$ can be employed to tune the system to a range of enhanced $S_N$ where the information processing capacity can be optimized.
The process of entanglement typically involves manipulating bipartite systems aiming to maximize the entangled state. However, as the number of qubits increases, maintaining and increasing entanglement in multipartite systems becomes increasingly challenging due to the fragility of entanglement, which can be easily disrupted by environmental noise and other sources of interference. Hence, our proposal for increasing $S_N$ upon tuning the system close to QCPs/QPTs varying specific tuning parameters might be relevant for the field of quantum computing. Such a method could simplify the process of entanglement control, potentially leading to more efficient and reliable quantum algorithms, as well as to more robust quantum hardware.

Besides the discussions regarding $\Gamma^{0\textmd{K}}$ as an ``\emph{entanglement compass}'', we now discuss the slowing down of the system dynamics and the “creation of mass” close to any QCP/QPT. Following Anderson's statement in his seminal paper ``\emph{The plasma frequency is equivalent to the mass}'' \cite{anderson} and based on our previous work \cite{griffiths}, we propose that symmetry breaking and fluctuations of the order parameter on the verge of any critical point/phase transition is inherently associated with the slowing down of the system's dynamics, which gives rise to low-energy excitation modes and emulates mass. This can be found, for instance, in the magnetic and electronic Griffiths phases \cite{vojta,griffiths}. In other words, as a direct consequence of the phases competition in the critical regime, the appearance of ``mass'' leads to a slowing down of the system's dynamic, which in turn reflects in an enhancement of both relaxation time \cite{griffiths} and effective mass \cite{senthil,paschen}. Such fascinating phenomena can be explored using $\Gamma^{0\textmd{K}}$ and $\Gamma$ \cite{griffiths,entanglement}.

In summary, we have proposed a new form of the Gr\"uneisen parameter valid at absolute zero temperature. The here-proposed $\Gamma^{0\textmd{K}}$ can be considered as an \emph{entanglement compass} to explore QPTs/QCPs upon varying tuning parameters. Considering that $\Gamma^{0\textmd{K}}$ quantifies the entanglement as a function of the tuning parameters, it can be employed to optimize information processing, in particular in the immediate vicinity of QCPs/QPTs. We have employed the 1D IMTF and Kane's quantum computer as working-horses to showcase our proposal. Furthermore, we have demonstrated that genuine QPTs/QCPs take place only when $E_0$ is non-linear with respect to the control parameter, which is key in quantum computing regarding both bipartite and multipartite systems. Also, we propose that in the thermodynamic limit a breakdown of the Hellmann-Feynmann theorem takes place at any QCP and we anticipate that this should also occur for the Bose-Einstein condensation \cite{entanglement}. It is challenging to put the present concepts in practice in real systems. Yet, the ``creation of mass'' in the vicinity of critical points/phase transitions was discussed.
\\

MdeS acknowledges financial support from the S\~ao Paulo Research Foundation Fapesp (Grants No. 2011/22050-4, 2017/07845-7, 2019/24696-0). National Council of Technological and Scientific Development CNPq (Grants No. \,302887/2020-2, 308695/2021-6) is acknowledged. This work was partially granted by Coordena\c c\~ao de Aperfeiçoamento de Pessoal de N\'ivel Superior - Brazil (Capes) - Finance Code 001 (Ph.D. fellowship of L.S.). LSR acknowledges the support from Icelandic Research Fund (Rannis), Projects No. 163082-051 and Hybrid Polaritonics.


\begin{thebibliography}{10}
\bibitem{stanley} H.E. Stanley, Introduction to Phase Transitions and Critical Phenomena (Oxford Science Publications, New York, 1971).
\bibitem{sethna} J.P. Sethna, Statistical mechanics: entropy, order parameters, and complexity - second edition (Oxford University Press, 2021).
\bibitem{zhu} L. Zhu, M. Garst, A. Rosch, Q. Si, Phys. Rev. Lett. \textbf{91}, 066404 (2003).
\bibitem{mrb} L. Squillante, Isys F. Mello, A.C. Seridonio, M. de Souza, Mat. Res. Bull. \textbf{142}, 111413 (2021).
\bibitem{Munoz2013} P. Moretti, M.A. Mu\~noz, \emph{Nat. Commun.} \textbf{4}, 2521 (2013).
\bibitem{grigolini} P. Grigolini, Chaos Solit. \textbf{81}, 575 (2015).
\bibitem{ejp} M. de Souza, P. Menegasso, R. Paupitz, A. Seridonio, R.E. Lagos,  Eur. J. Phys. \textbf{37}, 055105 (2016).
\bibitem{elastocaloric} L. Squillante, I.F. Mello, A.C. Seridonio, M. de Souza, Sci. Rep. \textbf{11}, 9431 (2021).
\bibitem{griffiths} Isys F. Mello, L. Squillante, G.O. Gomes, A.C. Seridonio, M. de Souza, J. Appl. Phys. \textbf{128}, 225102 (2020).
\bibitem{prl} L. Bartosch, M. de Souza, M. Lang, Phys. Rev. Lett. \textbf{104}, 245701 (2010).
\bibitem{unveiling} L. Squillante \emph{et al.}, %Isys F. Mello, G.O. Gomes, A.C. Seridonio, R.E. Lagos-Monaco, H. Eugene Stanley, M. de Souza,
Sci. Rep. \textbf{10}, 7981 (2020).
\bibitem{dobes} P. Cejnar, S. Heinze, J. Dobes, Phys. Rev. C \textbf{71}, 011304 (2005).
\bibitem{roscilde} L.-P. Henry \emph{et al.}, Phys. Rev. B \textbf{85}, 134427 (2012).
\bibitem{crossderivative} H.Y. Wu \emph{et al.}, arXiv:2210.08250v2 (2023).
\bibitem{neumann} J. von Neumann, Mathematical Foundations of Quantum Mechanics (Princeton University Press, Princeton, 1955).
\bibitem{sakurai} J.J. Sakurai, Modern Quantum Mechanics (Addison Wesley, 1994).
\bibitem{wootters1} W.K. Wootters, Phys. Rev. Lett. \textbf{80}, 2245 (1998).
\bibitem{fazio} A. Osterloh, L. Amico, G. Falci, R. Fazio, Nature \textbf{416}, 608 (2002).
\bibitem{wootters2} S.A. Hill, W.K. Wootters, Phys. Rev. Lett. \textbf{78}, 5022 (1997).
\bibitem{ney} A. Ney, T. Kammermeier, K. Ollefs, S. Ye, V. Ney, T. C. Kaspar, S. A. Chambers, F. Wilhelm, A. Rogalev, Phys. Rev. B \textbf{81}, 054420 (2010).
\bibitem{prb} G.O. Gomes \emph{et al.}%L. Squillante, A.C. Seridonio, A. Ney, R.E. Lagos, M. de Souza
, Phys. Rev. B \textbf{100}, 054446 (2019).
\bibitem{asadian} A. Montakhab, A. Asadian, Phys. Rev. A \textbf{82}, 062313 (2010).
\bibitem{feynman} R.P. Feynman, R.B. Leighton, M. Sands, The Feynman lectures on Physics - Quantum Mechanics (vol. 3), (Addison Wesley, 1965).
\bibitem{kondo} I. N. Karnaukhov, arXiv:2304.07233v1 (2023).
\bibitem{marra} P. Marra, J. Appl. Phys. \textbf{132}, 211101 (2022).
\bibitem{sachdev} S. Sachdev, Quantum phase transitions (Cambridge University Press, 2011).
\bibitem{pfeuty} P. Pfeuty, Ann. Phys. \textbf{57}, 79 (1970).
\bibitem{qsi} J. Wu, L. Zhu, Q. Si, J. Phys.: Conf. Ser. \textbf{273}, 012019 (2011).
\bibitem{bog} N.N. Bogoljubov, Il Nuovo Cimento \textbf{7}, 794 (1958).
\bibitem{soh} S. Oh, Eur. Phys. J. D \textbf{58}, 409 (2010).
\bibitem{sebastian} S. Fey, G. Kreimer, Investigation of zero-temperature transverse-field Ising models with long-range interactions, Ph.D. Thesis (2020).
\bibitem{gradshteyn} I.S. Gradshteyn, I.M. Ryzhik, Table of integrals, series, and products, (Academic Press, 2007).
\bibitem{albash} T. Albash, D.A. Lidar, Rev. Mod. Phys. \textbf{90}, 015002 (2018).
\bibitem{meyer} D.A. Meyer, N.R. Wallach, J. Math. Phys. \textbf{43}, 4273 (2002).
\bibitem{kitaev} G. Vidal, J.I. Latorre, E. Rico, A. Kitaev, Phys. Rev. Lett. \textbf{90}, 227902 (2003).
%\bibitem{montakhab} R. Radgohar, A. Montakhab, Phys. Rev. B \textbf{97}, 024434 (2018).
\bibitem{steane} A. Steane, Rep. Prog. Phys. \textmd{61}, 117 (1998).
\bibitem{assad} L.O. Conlon \emph{et al.}, Nat. Phys. \textbf{19}, 351 (2023).
\bibitem{kane} B.E. Kane, Nature \textbf{393}, 133 (1998).
\bibitem{buzek} P. $\check{\textmd{S}}$telmachovi$\check{\textmd{c}}$, V. Bu$\check{\textmd{z}}$ek, Phys. Rev. A \textbf{70}, 032313 (2004).
\bibitem{anderson} P.W. Anderson, Physical Review \textbf{130}, 439 (1963).
\bibitem{supercooled} Gabriel O. Gomes, H. Eugene Stanley, Mariano de Souza, Sci. Rep. \textbf{9}, 12006 (2019).
\bibitem{vojta} T. Vojta, J. Phys. A: Math. Gen. \textbf{39}, R143 (2006).
\bibitem{senthil} T. Senthil, Phys. Rev. B \textbf{78}, 045109 (2008).
\bibitem{paschen} S. Paschen, Q. Si, Nat. Rev. Mat. Phys. \textbf{3}, 9 (2021).
\bibitem{entanglement} M. de Souza \emph{et al.} (in preparation).
\end{thebibliography}
\end{document}